# Formation of a Black Hole in the Dark[*]

I. Félix Mirabel[1,2], Irapuan Rodrigues[1]

[1]Service d'Astrophysique – CEA-Saclay. 91191 Gif-sur-Yvette, France,
[2]Instituto de Astronomía y Física del Espacio/Conicet. Bs As, Argentina
E-mail: fmirabel@cea.fr.

**We show that the black hole in the x-ray binary Cygnus X-1 was formed in situ and did not receive an energetic trigger from a nearby supernova. The progenitor of the black hole had an initial mass greater than 40 solar masses and during the collapse to form the $\sim 10$ solar mass black hole of Cygnus X-1, the upper limit for the mass that could have been suddenly ejected is ~1 solar mass, much less than the mass ejected in a supernova. The observations suggest that high-mass stellar black holes may form promptly, when massive stars disappear silently.**

---

It is believed that stellar black holes can be formed in two different ways: Either the massive star collapses directly into a black hole without a supernova (SN) explosion, or an explosion occurs in a protoneutron star, but the energy is too low to completely unbind the stellar envelope, and a large fraction of it falls back onto the short-lived neutron star, leading to the delayed formation of a black hole (*1*). Recently, it has been found that the black hole x-ray binary GRO J1655-40 was launched far from its birth place by an energetic SN explosion (*2*) at a runaway speed (*3*) of 120 km s$^{-1}$. But so far there has been no observational evidence for a black hole formed without an energetic SN explosion.

Cygnus X-1 (*4*) is a well studied Galactic black hole candidate. The x-ray source was identified with a high mass binary system, where the radial velocity measurements indicate a compact object massive enough to be a black hole rather than a neutron star. Cygnus X-1 can be classified as a microquasar (*5*) with a persistent radio counterpart that has been resolved as a compact relativistic jet (*6*), which allows one to observe with high precision the motion of the x-ray binary in the sky.

Cygnus X-1 is moving as the association of massive stars Cygnus OB3 (Fig. 1). Despite the different observational techniques used to determine the proper motions of the high mass x-ray binary and the association of stars, the magnitudes are similar and both are directed to the Galactic plane. This supports the hypothesis that Cyg OB3 is the parent association of Cygnus X-1. Based on the proper motions and radial velocities presented in Table 1, at a distance of 2.0±0.1 kpc the velocity of Cygnus X-1 relative to Cyg OB3 is $9 \pm 2$ km s$^{-1}$, which is typical of the random velocities of stars in expanding associations (*7*). The proper motion of Cygnus X-1 relative to Cyg OB3 implies that the x-ray binary would have reached its projected distance of $\sim 60$ pc from the center of Cyg OB3 in $(7 \pm 2) \times 10^6$ years.

A lower limit for the initial mass of the progenitor of the black hole can be estimated by assuming that all massive stars of the parent stellar association, including the progenitor of Cygnus X-1, were formed over a short time span (*8*). The main–sequence star of higher mass found in Cyg OB3 is of spectral type O7 V and has a mass of 40 times the mass of the sun (M$_\odot$) (*8*). Because more massive stars evolve faster, the lower limit for the initial mass of the progenitor is (40 $\pm$ 5) M$_\odot$. The upper limit for the initial mass would be equivalent to that of the highest mass stars found in Galactic associations, up to $\sim$100 $M_\odot$. The time since the formation of Cyg OB3 and the progenitor of Cygnus X-1 as inferred from current models of stellar evolution (*9*) is $(5 \pm 1.5) \times 10^6$ years, which is -within the range of error- consistent with the $(7 \pm 2) \times 10^6$ years Cygnus X-1 would have taken to move from the center of Cyg OB3 to its present position.

Using the equations for symmetric mass ejection in black hole formation (*10*), we estimate the maximum mass that could have been suddenly ejected to accelerate the binary without disruption to a velocity of $(9 \pm 2)$ km s$^{-1}$. From the properties of Cygnus X-1 (Table 1) it is found that not more than (1$\pm$0.3) M$_\odot$ was ejected in the core collapse of the massive progenitor. Indeed, there is no observational evidence for a SN remnant in the radio continuum, x-ray, or atomic hydrogen surveys of the region where Cygnus X-1 was most likely formed.

Before collapse the progenitor must have lost $\geq$(30 $\pm$5) M$_\odot$, because the initial mass of the progenitor was $\geq (40\pm 5)M_\odot$, and the estimated black hole mass is (10 $\pm$5) M$_\odot$ (*11*). Some fraction of the missing mass may have been transfered to the binary companion, but because the later has a mass of $\sim$18 M$_\odot$, $\geq 12 M_\odot$ were lost by stellar winds. In such a case the progenitor of the black hole in Cygnus X-1 may have been a Wolf-Rayet star.

The formation of the black hole of Cygnus X-1 was not through a Type II SN, where hydrogen envelopes are blown away and the ejected masses are in the range of 10 to 50 M$_\odot$ (*12*), much greater than the upper limit for the mass that could have been suddenly ejected in Cygnus X-1. Alternatively, the core collapse could have occurred in a progenitor that lost its hydrogen-rich envelope (SN Ib), and even most of its helium envelope (SN Ic). Recent observations suggest that the energy and luminosity of an explosion in a SN of type Ib or Ic increase with an increasing amount of ejected mass (*12*), so the core-collapse onto the black hole in Cygnus X-1 was either underluminous with respect to typical supernovae, or occurred without an explosion. Thus stellar black holes, such as in Cygnus X-1, may form without a SN. If there is no SN associated with the formation of the black hole, then observers would not see an increase in energy or luminosity from the region. The black hole would form in the dark.

The maximum linear momentum and kinetic energy that could have been imparted to Cygnus X-1 by a SN trigger would be $(250 \pm 80)$ $M_\odot$ km s$^{-1}$ and $(2 \pm 0.5) \times 10^{46}$ erg, respectively. The maximum linear momentum for Cygnus X-1 is 2.5 times smaller than the linear momentum imparted by the SN (*2*) to the runaway black hole system GRO J1655-40. The upper limit for the runaway kinetic energy of Cygnus X-1 is at least 20 times smaller than that estimated (*3*) for GRO J1655-40, and $\sim 2 \times 10^{-5}$ that of a SN of $10^{51}$ ergs.

The kinematics of Cygnus X-1 and GRO J1655-40 suggest that the black holes in these two x-ray binaries were formed through different evolutionary paths. The black hole in GRO J1655-40 has a mass of $(5.4 \pm 0.3)$ $M_\odot$ (*13*) and was formed through an energetic SN explosion and fall-back on a neutron star. The black hole in Cygnus X-1 which has a mass of $(10.1 \pm 5)$ $M_\odot$ (*11*) was formed through a low energy explosion or even by prompt implosion without a SN. These observations are consistent with the theoretical model (*1*) in which the energy of the explosion in the core collapse of massive stars decreases as a function of the increasing mass of the progenitor and black hole. Although the observations discussed here are of black holes in x-ray binaries, dim (or dark) formation of black holes should also occur in massive progenitors that are not in binary systems, where the massive hydrogen envelope has been retained by the progenitor, and if there is a SN it will appear as a low-luminosity type II event.

The formation of Galactic black holes can be used as a local template to gain insight into the physics of the gamma-ray bursts of long duration, which are believed to come from relativistic jets produced during the formation of black holes in distant galaxies. The nature of the so called "dark gamma-ray bursts" i.e., those without x-ray and/or optical counterparts, has been intriguing. The optical and x-ray counterparts of gamma-ray bursts are afterglows produced by the shocks of the jets with circumstellar material composed by the stellar wind from the progenitor and/or the ejecta from the SN explosion. It has been proposed that gamma ray bursts may be dark because of dust obscuration, very high redshifts, or rapidly decaying transients. The analysis of observations reported here suggests that some gamma-ray bursts could be intrinsically dark. Indeed, it is known that the metallicity decreases with increasing redshift, and massive stars in the distant universe may produce weak stellar winds, and collapse promptly into high mass black holes, where there would be no massive stellar winds or SN ejecta to be shocked by the jets. In such a case, some dark gamma-ray bursts may be jets from massive stellar black holes formed in the dark, like the black hole in Cygnus X-1.

|  |  | **Cygnus X-1** |  | **Cyg OB3** |  |
|---|---|---|---|---|---|
| $l$ | [°] | 71.32 | (14) | 72.80 | (15) |
| $b$ | [°] | +3.09 | (14) | +2.00 | (15) |
| $\mu_\alpha$ | [mas yr$^{-1}$] | -4.2±0.2 | (6, 14) | -3.9±0.3 | (15) |
| $\mu_\delta$ | [mas yr$^{-1}$] | -7.6±0.2 | (6, 14) | -6.7±0.3 | (15) |
| V$_{helio}$ | [km s$^{-1}$] | -5.4±0.1 | (16) | -8.5±2.1 | (15) |
| D | [kpc] | 2.5±0.4 | (17) | 2.3±0.4 | (18) |
|  |  | $1.4^{+0.9}_{-0.4}$ | (14) | $1.8^{+0.7}_{-0.4}$ | (15) |
|  |  | 2.0 | (19) | 2.0±0.1 | (8) |
| M$_{BH}$ | [M$_\odot$] | 10.1±5 | (11) | — |  |
| M$_\star$ | [M$_\odot$] | 17.8±4.5 | (11) | — |  |
| Spect. type |  | O9.7Iab | (20) | — |  |

Table 1: **Data on Cygnus X-1 and Cyg OB3.** $l$ and $b$ are the Galactic longitude and latitude. The proper motion $\mu_\alpha$ and $\mu_\delta$ of Cygnus X-1 were determined by high-precision astrometry based on Very Long Baseline Interferometric (VLBI) observations of the compact radio counterpart in eight epochs between 1988 and 2001 (6, 14). The mean proper motions $\mu_\alpha$ and $\mu_\delta$ of Cyg OB3 were determined with the Hipparcos proper motions of 18 stars. V$_{helio}$ for Cygnus X-1 is the heliocentric radial velocity of the center of mass of the binary, and V$_{helio}$ of Cyg OB3 is the mean heliocentric radial velocity based on the radial velocities of 30 stars (15). D in kpc units is the distance estimated by different methods. M$_{BH}$, M$_\star$ and Spect. Type are the mass of the black hole, the mass of the donor star, and its spectral type, respectively. Bibliographic references are indicated in parentheses.

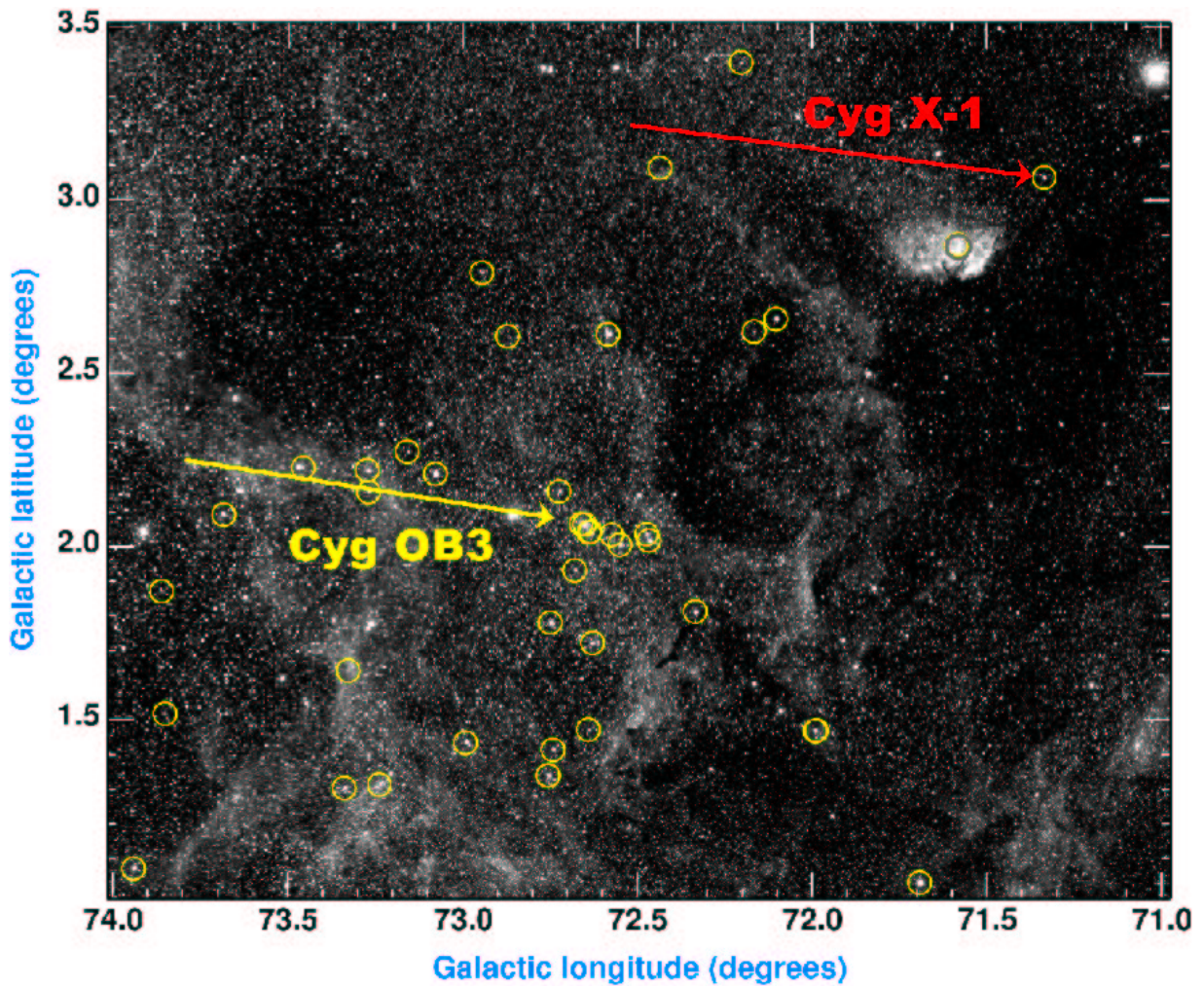

Figure 1: **Optical image of the sky around the black hole X-ray binary Cygnus X-1 and the association of massive stars Cyg OB3.** The red arrow shows the motion in the sky of the radio counterpart of Cygnus X-1 for the past 0.5 million years. The yellow arrow shows the average Hipparcos motion (*15*) of the massive stars of Cyg OB3 (*21, 22, circled in yellow*) for the past 0.5 million years. Despite the different observational techniques used to determine the proper motions, Cygnus X-1 moves in the sky as Cyg OB3. At a distance of 2 kpc the space velocity of Cygnus X-1 relative to that of Cyg OB3 is $9 \pm 2$ km s$^{-1}$.